\documentclass[iop]{emulateapj}
\usepackage{graphicx}
\usepackage{subfigure}
\usepackage{rotating} 

\newcommand{\lsim }{{\lower0.8ex\hbox{$\buildrel <\over\sim$}}}
\newcommand{\gsim }{{\lower0.8ex\hbox{$\buildrel >\over\sim$}}}

\def\HST{${\it HST}$}

\def\simge{\mathrel{%
  \rlap{\raise 0.511ex \hbox{$>$}}{\lower 0.511ex \hbox{$\sim$}}}}
\def\simle{\mathrel{
  \rlap{\raise 0.511ex \hbox{$<$}}{\lower 0.511ex \hbox{$\sim$}}}}

\newcommand{\Msun}{\ifmmode {M_{\odot}}\else${M_{\odot}}$\fi}
\newcommand{\Lsun}{\ifmmode {L_{\odot}}\else${L_{\odot}}$\fi}
\newcommand{\Rsun}{\ifmmode {R_{\odot}}\else${R_{\odot}}$\fi}

\usepackage{xargs}
\usepackage{color}

\shorttitle{Stellar Encounter Rate In Galactic Globular Clusters}
\shortauthors{Bahramian et al.}

\begin{document}
\title{STELLAR ENCOUNTER RATE IN GALACTIC GLOBULAR CLUSTERS}

\author{Arash~Bahramian\altaffilmark{1},  Craig~O.~Heinke\altaffilmark{1}, Gregory R. Sivakoff\altaffilmark{1}, Jeanette C. Gladstone\altaffilmark{1}}

\altaffiltext{1}{Dept. of Physics, University of Alberta, CCIS 4-183, Edmonton, AB, T5K 1V4, Canada; bahramia@ualberta.ca}

\begin{abstract}
The high stellar densities in the cores of globular clusters cause significant stellar interactions. These stellar interactions can produce close binary mass-transferring systems involving compact objects and their progeny, such as X-ray binaries and radio millisecond pulsars. Comparing the numbers of these systems and interaction rates in different clusters drives our understanding of how cluster parameters affect the production of close binaries. In this paper we estimate stellar encounter rates ($\Gamma$) for 124 Galactic globular clusters based on observational data as opposed to the methods previously employed, which assumed ``King-model" profiles for all clusters. By deprojecting cluster surface brightness profiles to estimate luminosity density profiles, we treat ``King-model'' and ``core-collapsed'' clusters in the same way.  
In addition, we use Monte-Carlo simulations to investigate the effects of uncertainties in various observational parameters (distance, reddening, surface brightness) on $\Gamma$, producing the first catalog of GC stellar encounter rates with estimated errors. Comparing our results with published observations of likely products of stellar interactions (numbers of X-ray binaries, numbers of radio millisecond pulsars, and $\gamma$-ray luminosity) we find both clear correlations and some differences with published results. 

\end{abstract}


\maketitle

\section{Introduction}
Soon after the discovery of bright X-ray binaries (XRBs) ($L_X\geq10^{34}$ erg/s) in our Galaxy, it became apparent that they were overabundant (by a factor of $\sim$100 per stellar mass) in globular clusters (GCs). This overabundance was attributed to the formation of XRBs by stellar interactions \citep{Clark75}.  Models of how neutron star XRBs could be produced dynamically include tidal capture of a companion star by a neutron star \citep{Fabian75}, collisions of neutron stars with giant stars \citep{Sutantyo75}, and exchange of neutron stars into existing primordial binaries \citep{Hills76}. These interactions depend on bringing two stars, or a star and a binary, close together, and thus depend on the square of the stellar density.  Gravitational focusing will bring stars closer together and is reduced by the stellar velocity dispersion, leading to a dependence of the stellar encounter rate (typically denoted $\Gamma$) on cluster properties as $\Gamma \propto \int \rho^{2}/\sigma$, where $\rho$ is the stellar density and $\sigma$ is the velocity dispersion.  

Globular cluster stellar distributions have often been found to be accurately described by lowered, truncated Maxwellian potentials, known as King models \citep{King62,King66}. These models possess a core region of nearly constant and a rapid falling off of density outside the core.  The majority of past work approximated the total $\Gamma$ of a cluster by only considering the summed $\Gamma$ within the core, assuming a constant density in the core; thus $\Gamma_1 \propto \rho^2 r_c^3/\sigma$, where $r_c$ is the physical radius of the cluster core.  Additional approximations based on King model profiles have been used, particularly when $\sigma$ is not well-known for a cluster. In a King model profile, $\sigma \propto \rho^{0.5} r_c$, so $\Gamma_2 \propto \rho^{1.5} r_c^2$ \citep{Verbunt87}. To date, even the most advanced calculations of $\Gamma$  that have integrated $\rho^{2}/\sigma$ have assumed the GCs follow a King model profile \citep[e.g.,][]{Pooley03}.  

These estimates have allowed comparison of the stellar interaction rates between different clusters in our galaxy, which showed that bright XRBs in Galactic globular clusters are indeed most concentrated in the highest-$\Gamma$ clusters \citep{Verbunt87, Verbunt02}.  Although it is more difficult to measure the surface brightness (SB) profiles of globular clusters in other galaxies, analysis of extragalactic globular cluster XRBs shows that they, too, tend to be concentrated in clusters that show evidence of higher $\Gamma$ values \citep{Jordan04,Sivakoff07, Jordan07, Peacock09}.  Evidence for a weaker-than-linear relation between $\Gamma$ and the probability of hosting a bright XRB in other galaxies (e.g., the nonlinear dependence of \citealt{Jordan04} can be explained by random errors in the measurements of cluster structural parameters, \citealt{Maccarone11}).  

In our own Galaxy, however, we have accurate radial SB measurements of globular clusters, allowing precise estimates of $\Gamma$.  The number of bright Galactic globular cluster XRBs is still too small for precise tests of stellar encounter theories.  However, recent X-ray, radio, and $\gamma$-ray observational advances provide large numbers of faint X-ray sources \citep{Pooley06}, radio millisecond pulsars \citep[MSPs,][]{Ransom08}, and integrated $\gamma$-ray emission that is presumed to arise from MSPs \citep{Abdo10}.  These results allow detailed comparisons between $\Gamma$ and the progeny of stellar encounters, X-ray binaries (both neutron star and white dwarf systems) and millisecond pulsars (the descendants of X-ray binaries).  

However, current literature calculations of globular cluster stellar encounter rates only approximate  the true density profile of the stellar cluster. The actual density profiles of many clusters do not exactly fit King models--there are 29 clusters in the Harris catalog \citep[][2010 edition; hereafter HC]{Harris96} with designations of ``core-collapsed'', or possibly core-collapsed.  Core-collapsed is an observational designation indicating that instead of showing a clear, flat central core, the radial SB profile of a cluster continues to increases towards its center.  These observations are linked to theoretical models of a gravitational instability that leads to a rapidly shrinking core \citep{Meylan97}, although the definition of core-collapse used by theorists does not necessarily coincide with the definition used by observers (compare \citealt{Hurley12} and \citealt{Chatterjee12}).
  In addition to core-collapsed clusters, many clusters that have generally been considered  to be well-fit by King models (e.g., NGC 6388) show radial SB gradients down to their centers, which are not predicted by King models \citep{Noyola06}.  Calculations of stellar encounter rates using different methods (e.g. $\Gamma_1$, \citealt{Heinke03d}; $\Gamma_2$, \citealt{Maxwell12}; integration of $\rho^2/\sigma$ of a King-model fit, \citealt{Pooley03}) can get significantly different results, implying that the choice of method introduces a systematic uncertainty.  This is a particular concern when considering how observationally core-collapsed clusters compare to other clusters, as none of the methods cited above use accurate descriptions of core-collapsed cluster properties (e.g., King-model fits to core-collapsed clusters simply assume a concentration parameter, $c$, of 2.5, which overestimates the SB gradient outside the core).

Moreover, previous calculations of stellar encounter rates have not, to our knowledge, quantified the uncertainties in their calculations. This makes it difficult to understand, when comparing  $\Gamma$ versus observations of close binaries, whether uncertainties in the input quantities, such as reddening, distance, or core radius (for $\Gamma_1$ or $\Gamma_2$), cause scatter in the correlations.  

Our goal in this paper is to rectify these two problems by calculating the 3-d radial luminosity density profile and integrating it to obtain an estimate of $\Gamma$. We then quantify the uncertainties in our calculations by Monte Carlo sampling from distributions of the observational inputs.  Finally, we compare our results with some recent works to determine how our estimates affect the correlation of stellar encounter progeny with stellar encounter rates. Note that the goal of this paper is simply to perform an accurate computation of the simplest stellar encounter rate estimate, and its errors.  We do not attempt here to include issues such as mass segregation, neutron star escape at birth, subsequent binary destruction, {}{dynamical evolution of GCs, finite lifetimes}, etc., which have been discussed in several works \citep[e.g.][]{Verbunt88,Verbunt03,Smits06,Ivanova08}, as they do not yet have simple, agreed-upon recipes that could be used to address these details.  We will model these effects in an upcoming paper, where we will draw further conclusions about the dynamics of XRB production.\\

\section{Data reduction and analysis}\label{sec:data}

To calculate $\Gamma$ based on  $\int \rho^{2}(r)/\sigma \, dr$ {over several half-light radii}, we need the luminosity density profile (as a function of radius), and velocity dispersions (ideally, also as a function of radius, but see below), along with estimates of the distance modulus and extinction. \\

\subsection{Surface Brightness Profiles}
Our sample includes 124 Galactic GCs for which found published SB profiles. For 85 GCs we used the SB profiles compiled by \citet[][hereafter T95]{Trager95}. These datasets were obtained from various
ground-based observations, mostly from the Berkeley Globular Cluster survey
by \citet{Djorgovski86}.  T95 indicate the quality of the datapoints with a {weight} and their best data are labeled with {weight}=1. 

\citet[][hereafter, NG06]{Noyola06} provide SB profiles for 38 GC, some of which are also listed in HC. In these overlapping cases, we use the SB profiles provided by NG06 as they were constructed from {\it Hubble Space Telescope}
(\HST) observations, which are much higher resolution than ground-based data and were processed with attention to reducing the influence of the brightest (giant) stars. 

\begin{figure*}
\begin{center}
\includegraphics[scale=0.30]{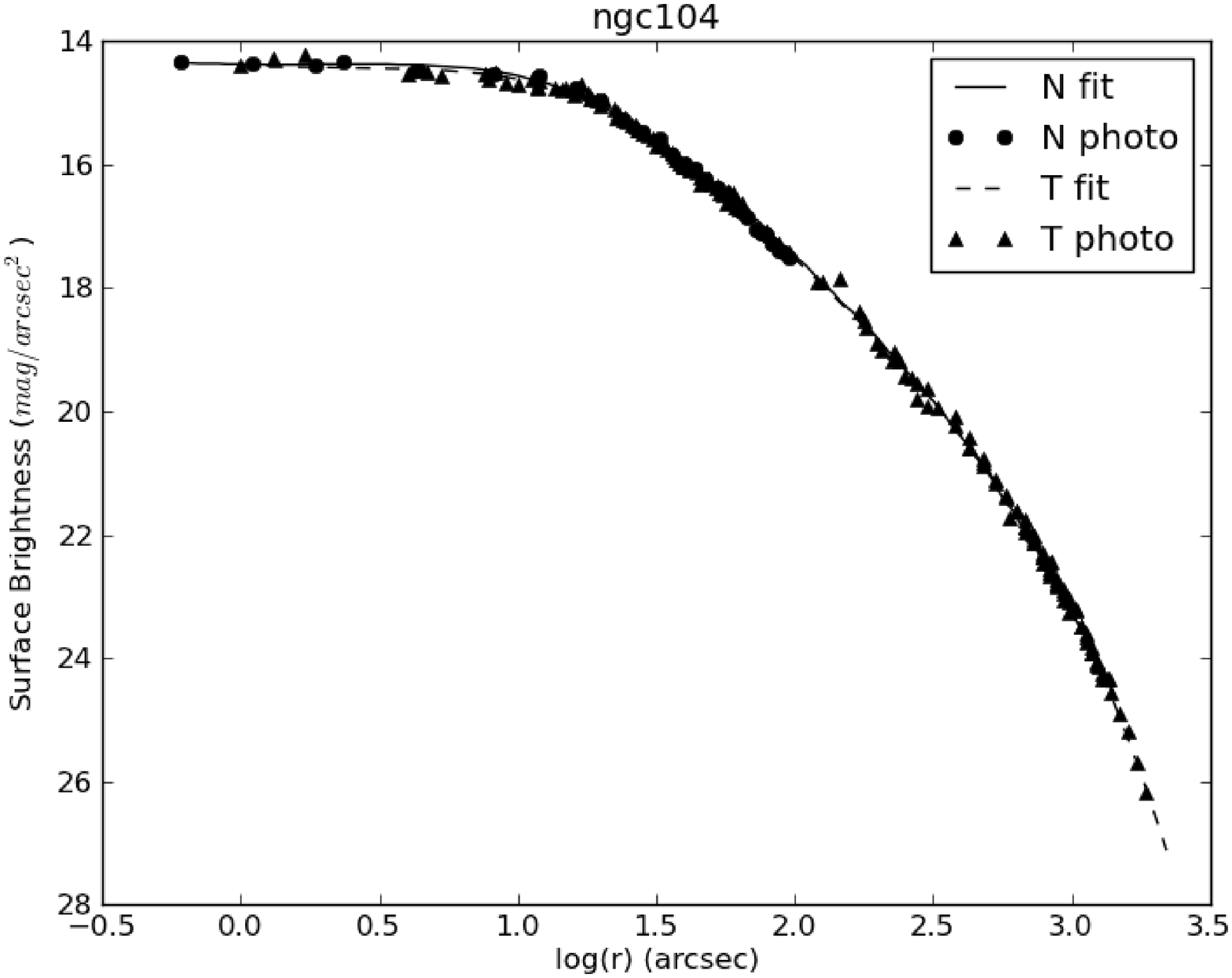}\hspace*{0.5cm}
\includegraphics[scale=0.30]{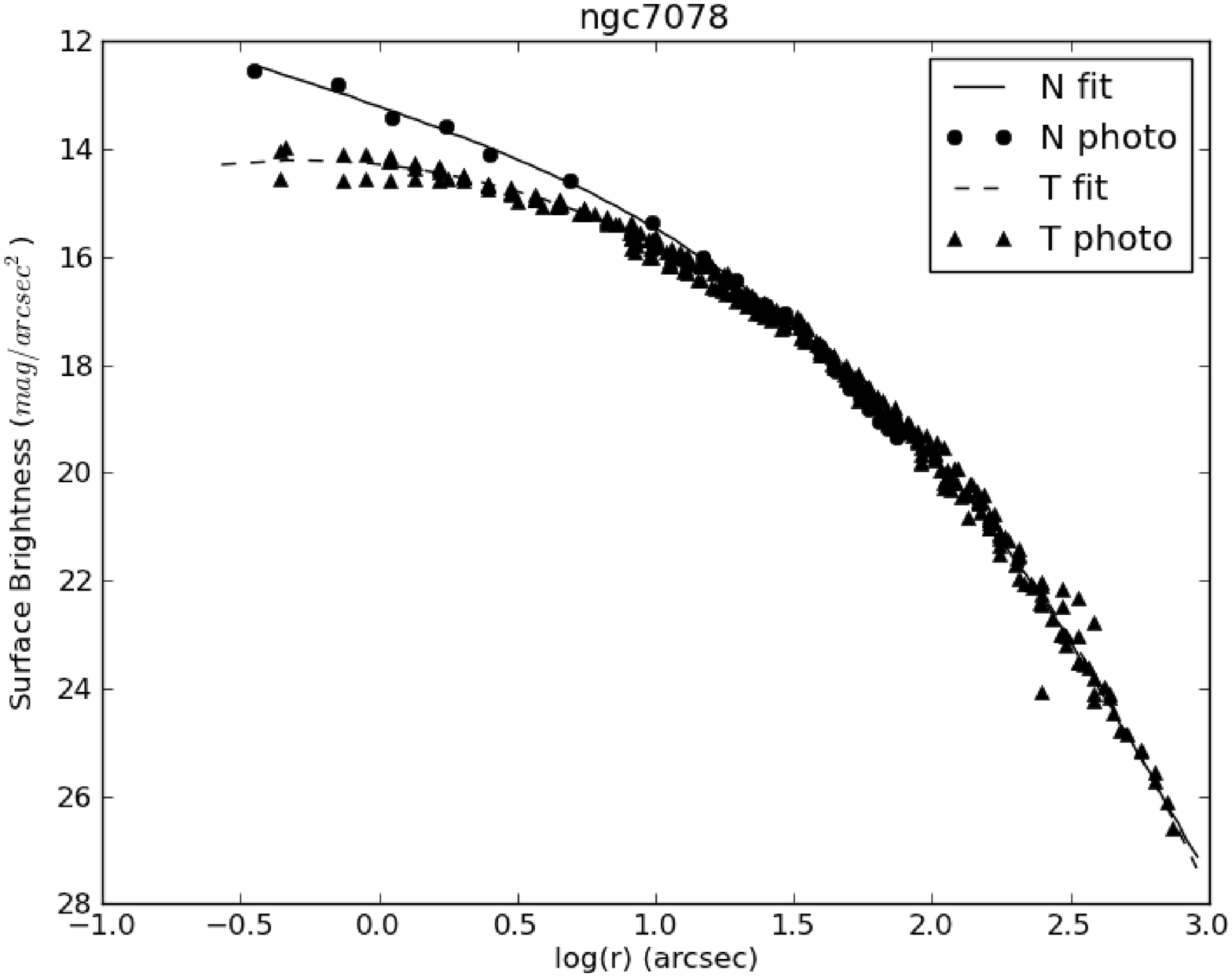}
\vspace*{0.5cm}
\includegraphics[scale=0.30]{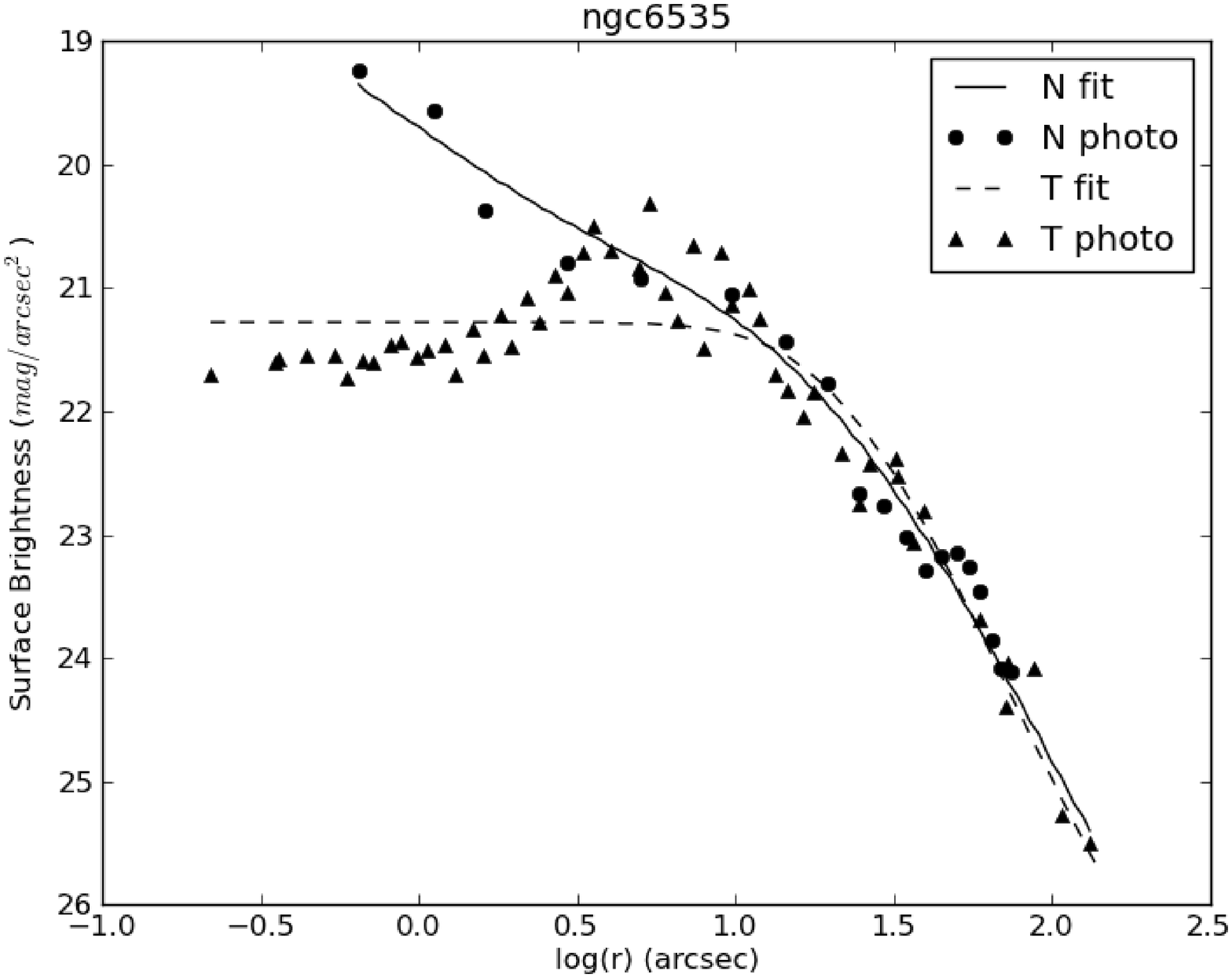}\hspace*{0.5cm}
\includegraphics[scale=0.30]{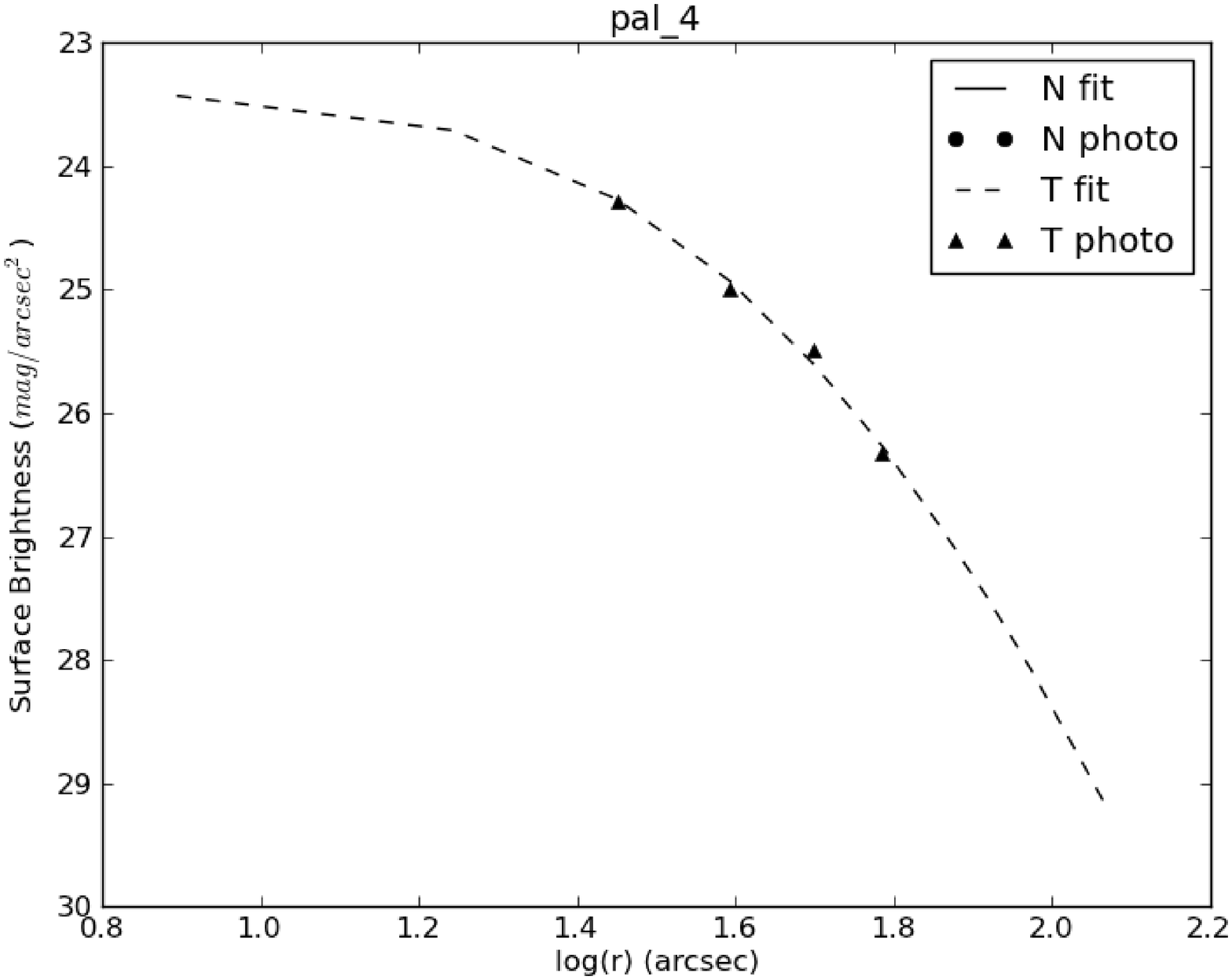}
\caption{Illustrations of SB profiles with different qualities. N fit is the fit provided by NG06. N photo are the photometric data points from NG06. T fit is the fit provided by T95. T photo are the photometric data points with weight=1.0 from T95.}
\label{sbp}
\end{center}
\end{figure*}

The quality of the observed SB data varies strongly from one GC to another (Fig.~\ref{sbp}). 
For all GCs except Terzan 5 (see details below), we used the Chebyshev polynomial fits provided in T95 or the spline fits provided by NG06, instead of the raw photometric data. Given both the noise in the SB profile data and the strong dependence of our method on the derivatives of the SB profiles, we used the smoothed profiles throughout this paper. As we show in \S \ref{gammares}, for GCs where the data is of high quality this choice has little effect on our calculations. For GCs with poor quality data, the Chebyshev polynomial fits lead to a smoother luminosity density profile that should be more representative of the actual luminosity density profile.

For three GCs (Palomar 10, Terzan 7, and Tonantzintla 2) the T95 SB profiles are
uncalibrated.  Following \citet{Mclaughlin05}, we calibrated these profiles 
by assuming that their central SB values are equal to the central SB values from the HC. For Terzan 7, in addition to calibrating the data, we ignored the polynomial fit data for $\log r_{\rm arcsec} >1.9$ to avoid the non-physical increase of the fit SB profile with radius.  Such a problem can be attributed to the lack of large-radius data points, and the high order of the Chebyshev polynomial fit. T95 also present two sets of data for NGC 2419. We choose the dataset which shows  agreement with the central SB reported by HC. 

We estimated uncertainties on the SB profiles using the reported uncertainties in the photometric data. 
For the NG06 SB profiles, we used the maximum reported uncertainty in photometric
data (requiring $\log r_{\rm arcsec} >0$). 
 For the T95 SB profiles, we used estimates of the photometric uncertainties calculated by \citet{Mclaughlin05}.

\begin{figure}
\includegraphics[scale=0.35]{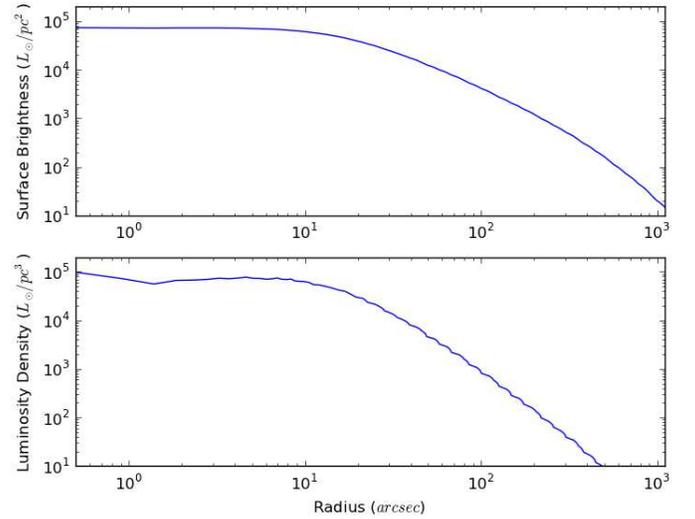}
\caption{Interpolated surface brightness profile (top, interpolated from values given by NG06) and deprojected luminosity density profile (bottom) for NGC 104.}
\label{sbplum}
\end{figure}

As all the profiles were reported as a function of angular radius, we first calculated 1-D profiles as a function of physical radius using the reported GC distances. To obtain 3-dimensional luminosity density profiles from the 1-dimensional observational SB profiles, we used the non-parametric deprojection of \citet{Gebhardt96} assuming spherical symmetry:
\begin{equation}\label{sbp_deprojection}
\rho(r) = -\frac{1}{\pi} \int_r ^\infty \frac{d \mu(R)}{d R} \frac{d
  R}{\sqrt{R^2 - r^2}}
\end{equation}
where $\mu(R)$ is the 1-D SB profile as a function of projected radius $(R)$
and $\rho(r)$ is the luminosity density as a function of deprojected
(spatial) radius $(r)$. When calculating the luminosity density function, we first linearly interpolated the (T95 and NG06) fits to the SB profile to allow for a finer numerical integration.
To integrate over the entire GC, we first set the central SB equal to the innermost data point (a very small extrapolation).  We then set the integration upper limit to be the outermost available data point which is in all cases $>2.5$ half-light radii, checking to ensure that this truncation did not affect our final results. In some cases where the SBD decreases inside the core (e.g., due to noise or contribution of light from giant stars outside the core), this integration yields a complex result. In all such cases, the imaginary component is less than 10$^{-6}$ the size of the real component. By ignoring this small imaginary component, we can reliably calculate the radial density distribution. Fig.~\ref{sbplum} shows the result of interpolation and deprojection for NGC 104, one of the most well-studied clusters.

\clearpage
\subsection{Velocity Dispersion}\label{sec:vel_disp}
\begin{figure}
\begin{center}
\includegraphics[scale=0.33]{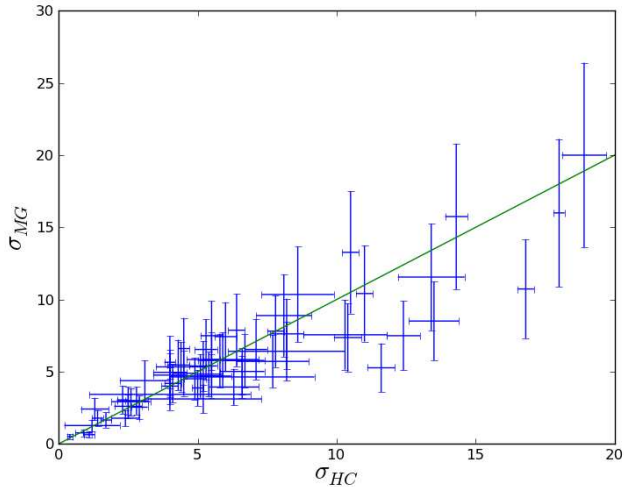}
\caption{Comparison of the modified central velocity dispersions from G02 ($\sigma_{\rm Modified\, Gnedin} = \sigma_{\rm Gnedin}/1.57$) to those from HC for 62 GCs in common (values in km/s). The line represents $\sigma_{HC}=\sigma_{MG}$.}
\label{sigma_compare}
\end{center}
\end{figure}

We have full velocity dispersion ($\sigma$) profiles for only 14 clusters {(see Table \ref{tab:sigma_nonflat} for these sources)}. For the remaining clusters, we only consider the central $\sigma$ value. Since $\sigma$ falls off much more slowly than SB with radius in the cluster, using the central $\sigma$ value for all radii produces very small changes in the inferred $\Gamma$ (see \S \ref{results}).
Our primary source for central values of $\sigma$ and their errors was HC, which compiles central velocity dispersion measurements for 62 GCs ($\sigma_{\rm HC}$).  For other GCs, we referred to theoretical estimates by \citet[][hereafter G02]{Gnedin02}.  For GCs where HC reports velocity dispersion, the G02 values are $1.57$ times larger on average. So for the cases where HC does not report velocity dispersion, we used modified values from G02, ($\sigma_{\rm MG} \equiv \sigma_{\rm G02}/1.57$). Fig.~\ref{sigma_compare} shows our comparison between the  $\sigma$ values from G02 (modified) and HC for the 62 clusters in common.

\begin{figure}
\begin{center}
\includegraphics[scale=0.33]{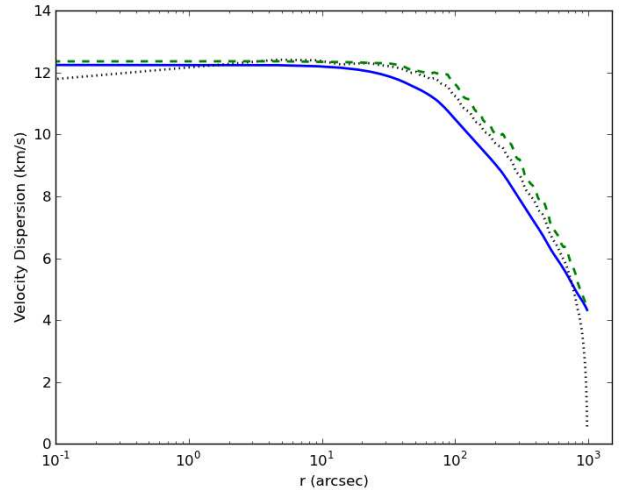}
\caption{Velocity dispersion profile for NGC 104. The solid line is the projected profile, the dashed line is the deprojected profile obtained from the sum, Eq.~\ref{sigma_deprojection2}, and the dotted line is the deprojected profile obtained from the integration, Eq.~\ref{sigma_deprojection1}. Note that the core radius for NGC 104 is $\approx21''$.}
\label{sigma_proj}
\end{center}
\end{figure}

For GCs where HC reports velocity dispersion, we used estimations he provides for uncertainty in the velocity dispersion. For the rest of our sample, we used the average fractional discrepancy between $\sigma_{\rm MG}$ and $\sigma_{\rm HC}$ for the 62 GCs they both report, as our uncertainty $\delta$:
\begin{equation}
\delta = \sqrt{\frac{1}{N}  \sum^N \left( \frac{\sigma_{\rm HC}-\sigma_{\rm MG}}{\sigma_{\rm HC}}\right)^2 } \approx 0.32
\label{eq:sigma_uncertainty}
\end{equation}
For the 14 GCs where we had detailed velocity dispersion profiles, we could compare the effects of assuming a constant velocity dispersion instead of using the true velocity dispersion profile. 
For these clusters, we deprojected the 1D velocity dispersion profile to a 3D profile making the assumption of spherical symmetry. We used the non-parametric integration for deprojection:
\begin{equation}\label{sigma_deprojection1}
\rho(r) \sigma(r) = -\frac{1}{\pi} \int_r ^\infty \frac{d (\mu(R) \sigma_{p}(R) )}{d R} \frac{dR}{\sqrt{R^2 - r^2}}
\end{equation}
where $\sigma_p(R)$ is the projected 1D profile and $\sigma(r)$ is
the deprojected 3D profile. Since the velocity dispersion data had not been previously smoothed, we applied a third-order interpolation prior to deprojecting the velocity dispersion. We truncated the integration at the outermost data point. This method produces a drop to zero at the outer radii, due to our choice of integration limits (choosing the outermost data point instead of infinity). 
To check the overall validity of the first method, we used a second method of deprojection assuming a discrete sum of shells, where we set  $\sigma_p$ and $\sigma$ to be equal in the outermost layer of the GC (we omit the factor of $\sqrt{3}$ in converting from 1-D to 3-D velocities, as it will be identical in all clusters, assuming isotropic orbits). By these assumptions, we calculate a discrete sum for the projection:
\begin{equation}\label{sigma_deprojection2}
\sigma_{p}(R_n) = \frac{\sum_i^n \rho(r_i) \sigma(r_i) }{\sum_i^n \rho(r_i)} 
\end{equation}
where $i$ starts from the outermost radius and goes towards the center. Starting from the outermost layer, we found values for the deprojected $\sigma$ at different points and interpolated them as a function of $r$. To compute  $\Gamma$ for these 14 GCs, we used the deprojected profile obtained from the latter method. In Fig.~\ref{sigma_proj} we present a comparison of the projected $\sigma_p$ profile and the deprojected $\sigma$ obtained from both methods for NGC 104.  In \S \ref{results} we discuss the effects on $\Gamma$ of using a full deprojected $\sigma$ profile versus assuming a constant $\sigma$ throughout the cluster.

\subsection{Distance Modulus and Extinction}
To estimate luminosity density as a function of physical radius for GCs, we need to calculate the physical radius using distance and angular radius. To calculate distance and estimate uncertainties on it, we used values for the apparent distance modulus $D(m-M)$ and foreground reddening $E(B-V)$ from HC. Based on the different claimed measurements in the literature for a few GCs, we assumed an uncertainty of $0.1$ magnitude in distance modulus for all GCs (except Terzan 5, see below). Following HC, we generally assumed
a $10~\%$ uncertainty for the reddening, $E(B-V)$, imposing a minimum uncertainty of $0.01$ magnitude for any cluster. We used $A_V = R_V E(B-V)$ with $R_V = 3.1$ to obtainthe extinction.  Since $R_V$ is not the same for all
parts of the sky \citep{Hendricks12,Nataf12}, we assumed a further uncertainty of $10~\%$ in $R_V$. For 3 GCs  (AM~1, NGC~5466, and NGC~7492) HC reports $E(B-V)=0.0$, in these cases we used alternative sources to improve these estimates. For AM~1 we chose 0.02 \citep{Dotter08}, for NGC~5466,
  0.02 \citep{Schlegel98}, and for NGC~7492, 0.04 \citep{Schlegel98}. 

\subsection{Special case of Terzan 5}

\begin{figure}
\begin{center}
\includegraphics[scale=0.33]{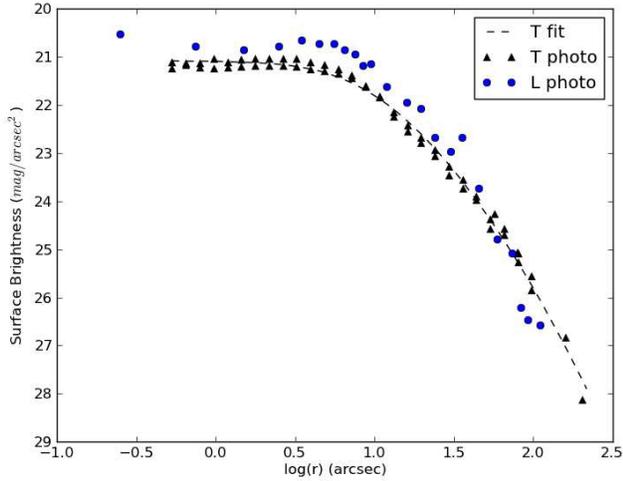}\hspace*{0.5cm}
\caption{Available SB data for Terzan 5. T fit:  fit provided by T95. T photo: photometric data points with weight=1.0 from T95. L photo: Photometric data from L10. In comparison L10 shows a higher SB value for the core.}
\label{sbp_terzan5}
\end{center}
\end{figure}

Terzan 5 is a highly extincted GC near the Galactic core that contains $>50$ XRBs \citep{Heinke06b} and $>33$ millisecond radio pulsars \citep{Ransom05}. This large population of sources makes it an ideal GC for more detailed analysis. Although SB profiles are available in T95, we note that higher quality data was available in \citet[][hereafter L10]{Lanzoni10}, derived using \HST observations (ACS - F606W). However, L10 did not provide clear fit parameters. As a result, we use their photometric data to derive SB (Fig.~\ref{sbp_terzan5}). We assume an uncertainty of 0.2 magnitudes for the SB profile, as reported by L10. Recently \citet{Massari12} presented a high resolution reddening map of Terzan 5.  From their map, we find $E(B-V) = 2.61$ for the core of Terzan 5 and used their estimate of $R_V=2.83$ to obtain our $A_V$ estimate. For its distance modulus we used the value of $(m-M)_v$=21.27 from HC, which with our $A_V$ gives the same $(m-M)_0$=13.87 as \citet{Valenti07} derive. However, due to the uncertainty in measuring $(m-M)_0$ in this highly reddened case, we assumed a conservative uncertainty of $0.2$ for this quantity. 

\section{Stellar encounter rate, $\Gamma$}\label{gammares}
To calculate $\Gamma$, we numerically integrated $a \rho(r)^{2}/\sigma(r)$ using the luminosity density and velocity dispersion profiles derived above, where $a$ is an arbitrary constant that was set by requiring the $\Gamma$ value for NGC~104 be equal to 1000.
To ensure that the first-order interpolation of the fits to the SB profile were appropriate, we recalculated $\Gamma$  using both second-order and third-order interpolation. This led to no significant differences in the final results ($\lesssim 0.1~\%$
  change). 

To estimate the uncertainty
in $\Gamma$, we performed Monte-Carlo simulations of the $\Gamma$
calculation with different inputs. Our principal code is written in {\it Mathematica}\footnote{http://www.wolfram.com} and the average number of iterations for each GC was
$\approx400$. We assumed gaussian distributions for the input parameters
(distance modulus, reddening, $R_V$, SB profile amplitude, velocity dispersion)
with the reported values as the mean value, and the reported uncertainties
as the standard deviations of the distributions. We used these distributions
with caution, modifying them when they were unphysical. For low values of extinction, the gaussian distributions include negative values.  For the velocity dispersion, values very near to zero also produce unphysical results (since velocity dispersion is in the denominator in $\Gamma\sim\int \rho/\sigma$). So we did not run simulation for those values.

In the case of extinction,
we required positive values, and in the case of velocity
dispersion,
we required that the simulated velocity dispersion was within two standard deviations (eq.\ref{eq:sigma_uncertainty}) of the measured velocity dispersion. For 2 GCs, NGC~7492 and NGC~5946, the reported
  uncertainties from HC on $\sigma$ are more than $50~\%$, so
  for these two, we truncated the $\sigma$ distribution at one standard deviation
  instead.

When the photometric data was of high quality, we found that integrating this data directly  gave similar results as integrating the fitted Chebyshev polynomials. 
The differences in the final results were typically $<1~\%$ (e.g. NGC 104). In Table \ref{tab:fit_photo} we provide  a comparison between $\Gamma$ calculated based on the photometric data,  and based on the Chebyshev fit for some of the GCs where  data were available from NG06. In the few cases with a large difference between the two values (e.g., NGC 5897, NGC 6205 \& NGC 6254), the observational data did not extend out to the outer portions of the GC. In these cases, by truncating the Chebyshev fit profile to the outermost point of the photometric data, we greatly reduce the difference in results; for NGC 5897 it drops to $26.7~\%$ and for NGC~6254 to $13.2~\%$.

\begin{table}
\begin{center}
\begin{tabular}{lccc}
\hline
GC & $\Gamma_{photometric}$ & $\Gamma_{fit}$ & difference~(\%)\\
\hline
NGC 104 & 992.6 & 1000 & 0.7\\
NGC 1851 & 1637 & 1528 & 7.1\\
NGC 1904 & 115.6 & 115.7 & 	0.9\\
NGC 2298 & 4.091 & 4.314	 & 5.2\\
NGC 2808 & 882.8 & 922.9 & 4.3\\
NGC 5272 & 172.4 & 194.4 & 11.3\\
NGC 5286 & 449.0 & 458.0 & 1.9\\
NGC 5694 & 207.1 & 191.1 & 8.3\\
NGC 5824 & 1046.4 & 984.3 & 6.3\\
NGC 5897 & 0.2845 & 0.850 & 66.5*\\
NGC 5904 & 152.42 & 164.1 & 7.1\\
NGC 6093 & 568.24 & 531.6 & 6.9\\
NGC 6205 & 48.475 & 68.91 & 29.6*\\
NGC 6254 & 13.656 & 31.37 & 56.5*\\
NGC 6266 & 1827.1 & 1666.5 & 9.6\\
NGC 6284 & 670.77 & 665.54 & 0.8\\
\hline
\end{tabular}
\end{center}
\caption{Comparison between $\Gamma$ calculated based on photometric data and based on the Chebyshev fits to the SB profiles (from NG06). Incompleteness in the photometric data appears to explain the cases with a large difference between the two $\Gamma$ values (marked by a *). In these cases, truncating the fit to the region where photometric data is available reduces the difference in results.}
\label{tab:fit_photo}
\end{table}

\section{Results}\label{results}
\begin{figure}
\includegraphics[scale=0.33]{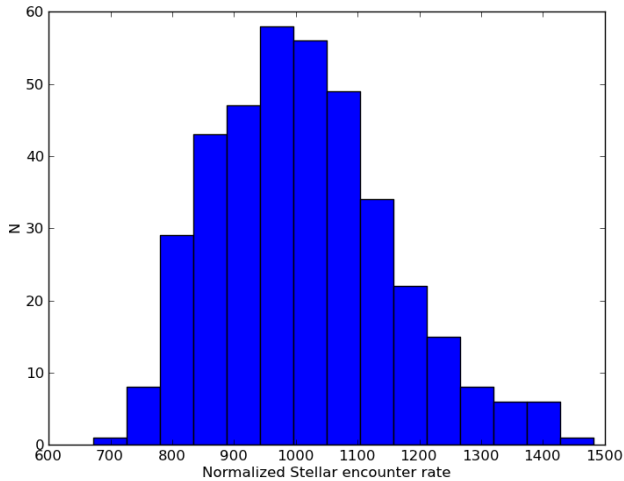}
\caption{Results of Monte-Carlo simulations for NGC 104, showing the number of trials giving each value for $\Gamma$.  Due to our choice of normalization, the histogram is forced to be centered on 1000.}
\label{histo}
\end{figure}

The final $\Gamma$ values we report (Table \ref{tab:gamma}) are calculated based
on the default values for quantities described in \S \ref{sec:data}.  For most clusters, the $\Gamma$ calculated from the default values lies within 5\% of the median of the histogram of $\Gamma$ values produced in our simulations (Generally the discrepancy between default value and median of the distribution is caused by truncation of the input parameters distribution described in \S\ref{gammares}). Uncertainties in $\Gamma$ for each source are calculated based on the histograms of $\Gamma$ values produced 
from our Monte-Carlo simulations (Fig.~\ref{histo}).  We identify the 1-$\sigma$ upper bound by increasing $\Gamma$ from the median of  the distribution upwards until we include an additional 34\% of the simulations, and similarly identify the 1-$\sigma$ lower bound.  (Note that the $\Gamma$ probability distribution is not necessarily a Gaussian.)

\begin{figure}
\begin{center}
\includegraphics[scale=0.45]{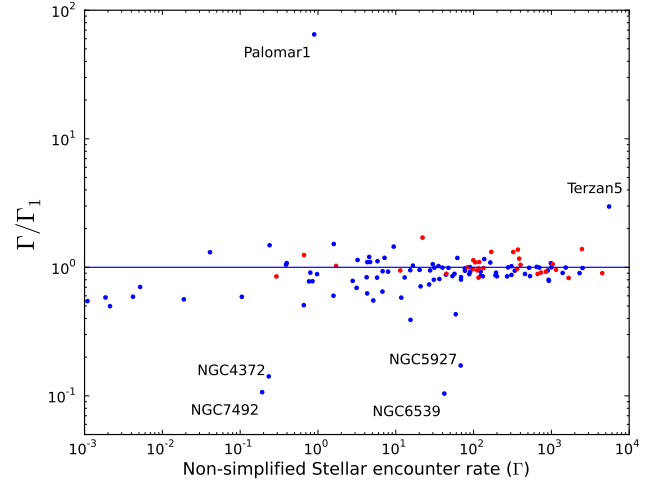}
\caption{Comparing $\Gamma_1 = \rho_c^2 r_c^3 \sigma_c^{-1}$ and $\Gamma = \sigma_c^{-1} \int \rho^2(r) d^3r $. Core-Collapsed GCs are denoted in red.}
\label{gamma_compare}
\end{center}
\end{figure}
  
\begin{table}
\begin{center}
\begin{tabular}{lccc}
\hline
Name & $\Gamma$ & Lower bound & Upper bound \\
\hline
Terzan 5   & 6800	& 3780	&7840 \\
NGC 7078 &4510	& 3520	&5870 \\
NGC 6715 &2520	& 2250	&2750 \\
Terzan 6   & 2470	& 753 	&7540 \\
NGC 6441 &2300 	& 1660	&3270 \\
NGC 6266 &1670 	& 1100	&2380 \\
NGC 1851 &1530 	& 1340	&1730 \\
NGC 6440 &1400 	& 923	&2030 \\
NGC 6624 &1150 	& 972	&1260 \\
NGC 6681 &1040 	& 848	&1310 \\
NGC 104   &1000 	& 866	&1150 \\
\hline
\end{tabular}
\end{center}
\caption{The 11 GCs with the highest $\Gamma$ values, providing their $\Gamma$ values (normalized to give NGC 104's $\Gamma$=1000), and 1-$\sigma$ bounds.  The complete table is available in the electronic edition of the journal.}
\label{tab:gamma}
\end{table}

We also investigated the effects of assuming a constant velocity dispersion
profile by comparing the $\Gamma$ computed based on a constant 
$\sigma$ profile versus the actual measured (and deprojected) $\sigma$ profile for 14
GCs.
 For the purposes of this comparison alone, we used the central
velocity dispersion values reported by these profiles as the value for the constant velocity dispersion calculations (instead of the values from HC or G02). For deprojecting the observed velocity dispersion profiles we used the method of sums described in Section \ref{sec:vel_disp}.  As shown in Table \ref{tab:sigma_nonflat}, the difference between the results is always less than 15\%, and usually less than 5\%. For the final values, for consistency, we used a constant $\sigma$ for all GC for the calculations presented in Table~\ref{tab:gamma} and \ref{tab:gamma_full}.

\begin{table}
\begin{center}
\begin{tabular}{lcc}
\hline
name & Difference~($\%$) & Ref. \\
\hline
NGC 104 &	1.13 & (1)\\
NGC 288 &	2.86 & (1)\\
NGC 362 &	2.63 & (1)\\
NGC 2419 & 2.94 & (1)\\
NGC 3201 & 5.00 &(1)\\
NGC 5024 & 0.23 & (2)\\
NGC 5139 & 1.32 & (1)\\
NGC 6121 & 0.08 & (1)\\
NGC 6218 & 14.7 & (1)\\
NGC 6254 & 0.15 & (1)\\
NGC 6341 & 2.56 & (1)\\
NGC 6656 & 4.05 & (1)\\
NGC 6809 & 7.89 & (1)\\
NGC 7078 & 2.84 & (3)\\
\hline
\end{tabular}
\end{center}
\caption{Difference between $\Gamma$ calculated based on a constant velocity dispersion and the measured velocity dispersion profile. References for velocity dispersion profiles - (1): \citealt{Zocchi12} (using their King model fits to the profiles), (2):\citealt{Sollima12}, (3):\citealt{Murphy12}}
\label{tab:sigma_nonflat}
\end{table}

To have a complete set of calculations, we also included $\Gamma$ calculations and uncertainty estimations based on the simplified equations ($\rho_c^2 r_c^3 \sigma_c^{-1}$ and $\rho_c^{1.5} r_c^2 $) for 143 GCs (19 GCs  in addition to the main sample) using HC (Table~\ref{tab:gamma_full}). To do this, we used central surface brightness (in magnitude per arcsec$^2$), $\mu$, extinction, distance modulus, estimated core radius, $r_c$ and concentration parameter, $c$, to calculate central luminosity density, $\rho_c$. Following the prescription from \citet{Djorgovski93}: 
\begin{equation}
\rho=\frac{10^{0.4(26.362-\mu)}}{p~r_c}
\end{equation}
where $p= 10^{-0.603\times10^{-c}+0.302}$ and $r_c$ is in parsec.

For velocity dispersion we used the central values that we aggregated from the literature in \S \ref{sec:vel_disp}. Similar to the method described in \S\ref{gammares} we did Monte-Carlo simulations to estimate their effects on $\Gamma$. We assumed uncertainties on extinction, distance modulus and surface brightness as before. We also assumed an uncertainty of $5\%$  for the core radius. Since the concentration parameter $c$ has little effect, we did not include any error on $c$.
Comparing these simplified values of $\Gamma$ to our main results, the differences are relatively small for many GCs (Fig.~\ref{gamma_compare}). Although the value of $\Gamma$ for Terzan 10 calculated by the simplified method ($\Gamma_2$) is extremely high, we found it to be untrustworthy. While HC reports the core radius of Terzan 10 is $\sim0.9'$, inspection of a 2-MASS J-band image from the Infrared Science Archive\footnote{\url{http://irsa.ipac.caltech.edu/}} suggests it is $<0.2'$.

\section{Applications}
\subsection{X-ray sources}
\begin{figure}
\begin{center}
\includegraphics[scale=0.45]{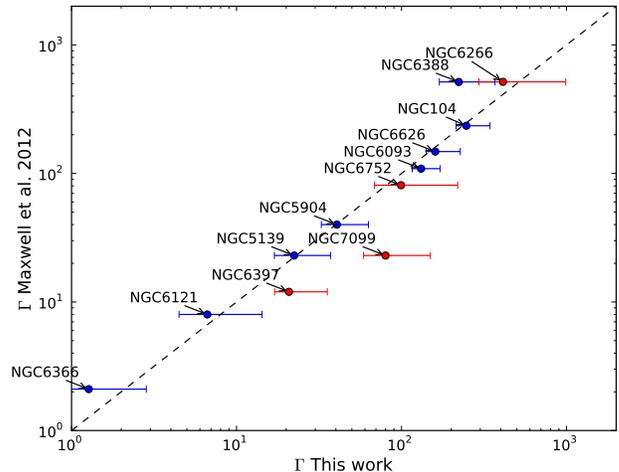}
\caption{Comparison of $\Gamma$ estimates by Maxwell et al. 2012 versus values calculated in this work (using a different normalization).  Core-collapsed clusters have errors shown in red, and show many of the largest differences. Note also that NGC 6388 has a lower $\Gamma$ in our calculations. We choose our $\Gamma$ normalization to give average values similar to those of Maxwell et al. 2012.}
\label{maxwell}
\end{center}
\end{figure}

A significant difference between our results and previous works comes in
the case of core-collapsed clusters. For instance, \citet{Maxwell12} derives similar values for $\Gamma_2$, with differences principally arising in the core-collapsed clusters (Fig. \ref{maxwell}). 

\begin{figure}
\begin{center}
\includegraphics[scale=0.45]{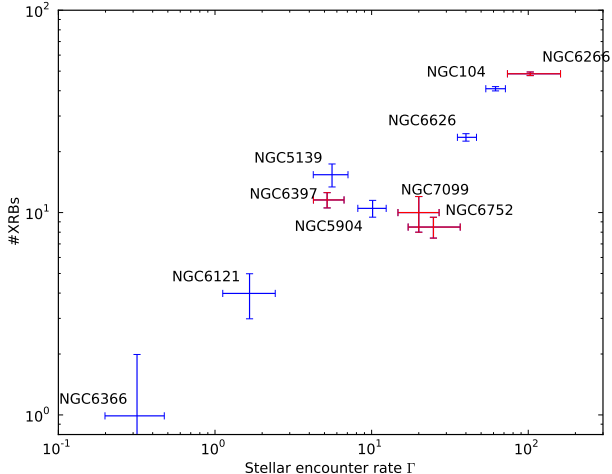}
\caption{Numbers of XRBs (from Pooley et al. 2003, Lugger et al. 2007), excluding the GCs with only lower limits determined), compared to our $\Gamma$ estimates, with appropriate error bars.  Core-collapsed clusters are in red, and show a tendency towards fewer XRBs for their $\Gamma$ than other clusters. We choose a $\Gamma$ normalization assuming $\Gamma$=20 for NGC 7099.}
\label{pooley}
\end{center}
\end{figure}

Comparing our values for $\Gamma$ to results from \citet{Pooley03} (which calculate $\Gamma$ by integration over the half-mass radius assuming king models) and \citet{Fregeau08}, our calculations show that, at about the same values of $\Gamma$, core-collapsed GCs have lower numbers of X-ray sources compared to typical GCs (Fig. \ref{pooley}). This is in contrast with the results of \citet{Fregeau08}. \citet{Fregeau08} suggested that, contrary to previous thinking, most globular clusters are currently still in their ``early'' contraction phase, and that only those clusters observationally defined as ``core-collapsed'' have reached the binary-burning phase. These clusters would then need to be currently ``burning'' binaries to support themselves at their current core radius.  The initial impetus for this suggestion was the apparent excess of X-ray sources in three ``core-collapsed'' clusters, NGC 6397, M30, and Terzan 1, compared to other GCs with similar values of $\Gamma$.  This would be explained if X-ray binaries were created a few Gyrs ago, at a time when non-core-collapsed clusters were substantially larger and less dense, but core-collapsed clusters presumably were at their current size.  Thus, the $\Gamma$ relevant for producing the current X-ray sources in non-core-collapsed clusters would be smaller than the currently observed $\Gamma$, as those clusters will have contracted and become denser. Our calculations remove the evidence for NGC 6397 and M30 having higher-than average X-ray source numbers for their  $\Gamma$ values. Instead our results suggest that core-collapsed clusters underproduce X-ray binaries.

One cluster that may not fit with this picture is Terzan 1. This is a GC that appears to be core collapsed, but its structural parameters are poorly determined at present. However, its position near the Galactic centre suggests an alternative scenario, that it may have been tidally stripped \citep{Cackett06a}. 
 
Fig.~\ref{pooley} indicates that, although the assertion about globular cluster evolution by \citet{Fregeau08} may or may not be true, the numbers of X-ray sources above $4\times10^{30}$ erg~s$^{-1}$ do not provide evidence for this assertion.  Other evidence, perhaps from comparing detailed Monte Carlo models of gravitational interactions between stars with observed quantities \cite[e.g.,][]{Chatterjee12}, may illuminate this question.  On the other hand, the X-ray sources in core-collapsed clusters will experience substantial binary destruction \citep{Verbunt03}, which may explain their rather different luminosity functions \citep{Pooley02b,Heinke03d,Stacey12}.

\subsection{Numbers of radio MSPs}
Large numbers of radio MSPs have been detected in several GCs, with the largest numbers in very high-$\Gamma$ clusters \citep{Camilo05}.  Several works have attempted to compare the numbers of MSPs in different clusters, accounting for the detection limits of the surveys of each cluster, to determine how cluster properties relate to MSP numbers \citep{Johnston92,Hessels07,Ransom08,Hui10,Lynch11,Bagchi11}.  These analyses must estimate the radio luminosity functions of cluster MSPs and the sensitivity of different surveys (involving complex estimates of pulsar detectability).  Perhaps the most sophisticated of these is that of \citet{Bagchi11}, which incorporates information from diffuse radio flux measurements \citep{Fruchter00,McConnell04} and summed $\gamma$-ray emission (e.g. \citealt{Abdo10,Hui11}; see also below).

\citet{Bagchi11} calculate the most likely numbers of MSPs in 10 globular clusters, based on their simulations of the detectability of MSPs in these clusters, and from the observations discussed above.  They make the striking claim that there is no compelling evidence for any direct relationship between any GC parameter and the number of MSPs per cluster; in particular, they claim that there is no correlation between $\Gamma$ and the number of MSPs.  \citet{Bagchi11} use Pearson, Spearman, and Kendall statistical correlation tests, and report the relevant coefficients and null-hypothesis probabilities.  We note that the null-hypothesis probabilities for the Spearman and Kendall tests for correlation between their calculated $\Gamma$ and the numbers of MSPs are 0.02 and 0.01, rather less than the typical 0.05 criterion for significance.  However, the Pearson test's null-hypothesis probability is only 0.07, which does not provide clear evidence of correlation.  

\begin{figure}
\begin{center}
\includegraphics[scale=0.45]{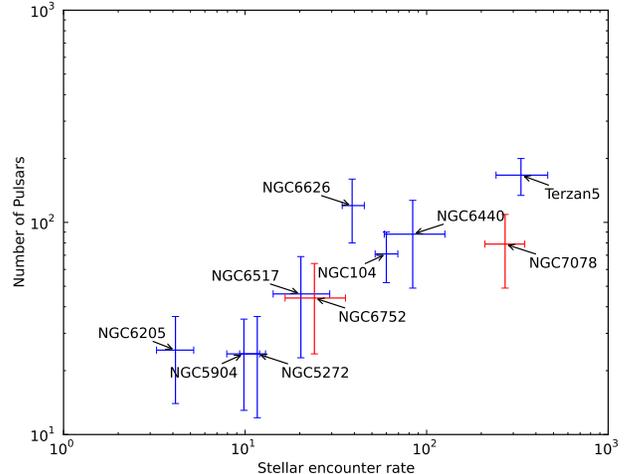}
\caption{Number of recycled pulsars within a GC (from Bagchi et al. 2011) compared to our calculated $\Gamma$ values. Core-collapsed clusters are shown in red. A correlation is clearly seen, and tabulated in Table \ref{tab:corr}.  The normalization of $\Gamma$ is chosen to be similar to the choice of normalization in Bagchi et al. 2011 ($\Gamma_{\rm NGC~6266} = 100$).}
\label{bagchi}
\end{center}
\end{figure}

Here we assume that their calculations of the numbers of MSPs are correct, and recalculate these correlations using our new $\Gamma$ values.  We use model 1 {(FK06)} from \citet{Bagchi11} for comparisons, as do they.
In Fig.~\ref{bagchi} and Table \ref{tab:corr}, we show and calculate the correlations between our values for $\Gamma$ and their MSP population results. 
Our statistical correlation tests indicate a very strong correlation between $\Gamma$ and the number of MSPs in a GC, with null-hypothesis probabilities of no correlation below $0.013$.  

\subsection{$\gamma$-ray fluxes}
The Fermi $\gamma$-ray Space Telescope's Large Area Telescope's unprecedented sensitivity and spatial resolution to GeV $\gamma$-rays have allowed detection of numerous radio MSPs as $\gamma$-ray sources \citep{Abdo09a, Abdo09b}, showing characteristic hard GeV spectra with cutoffs around 1-3 GeV \citep{Abdo09b}.  Fermi has recently detected 
gamma-ray emission from several globular clusters, including 47 Tuc and Terzan 5  \citep{Abdo09c,Kong10,Abdo10}, showing similar $\gamma$-ray spectra as radio MSPs, indicating that the observed $\gamma$-ray flux is likely due to a population of $\gamma$-ray-emitting MSPs.  In many clusters, no periodicities have been identified in the $\gamma$-ray emission, indicating that numerous MSPs contribute to the total emission, and thus that measurements of the total $\gamma$-ray flux can be used to estimate the number of MSPs in the cluster.  However, NGC 6624 shows a counter-example, where a single MSP dominates the $\gamma$-ray flux \citep{Freire11}, indicating that this method of estimating MSP numbers has limitations.  Recent claims of detections of $\gamma$-ray fluxes from globular clusters have been made for $\gamma$-ray sources lying well outside the half-mass radius of clusters, at low significance, and without evidence of spectral similarity to radio MSPs \citep{Tam11}.  We do not trust that these $\gamma$-ray sources represent the MSP population of these GCs and therefore choose to evaluate the effects of our calculations of $\Gamma$ on the correlations between integrated $\gamma$-ray flux and $\Gamma$ discussed by \citet{Abdo10}.

\begin{figure}
\begin{center}
\includegraphics[scale=0.46]{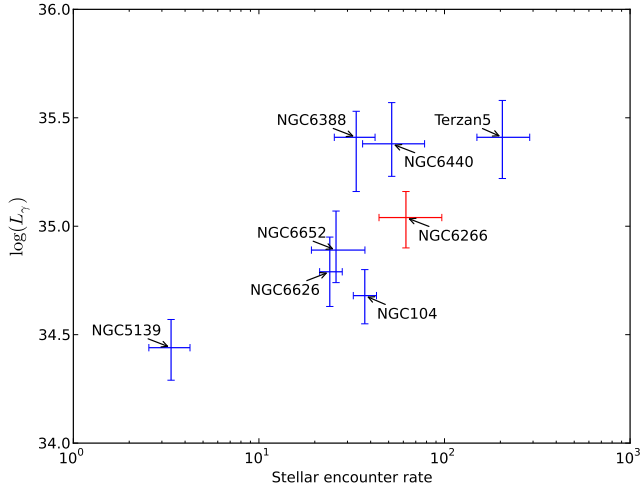}
\caption{Gamma-ray flux measurements from Abdo et al. 2010 ($\gamma$) versus our calculated $\Gamma$ values ($\Gamma$). Core-collapsed clusters are shown in red. A correlation is suggested, and tabulated in Table \ref{tab:corr}. The normalization of $\Gamma$ is scaled so that $\Gamma_{\rm NGC~6121} = 1$.}
\label{abdo}
\end{center}
\end{figure}

\citet{Abdo10} measured $\gamma$-ray luminosities and calculated 
$\Gamma$ for 8 GCs to investigate for a correlation. Using
their reported values for $\gamma$-ray luminosities and our estimates
for $\Gamma$, we find evidence (i.e., the probability that such a correlation occurs randomly is less than 10~$\%$) for a correlation between the two
parameters (Fig. \ref{abdo}, Table \ref{tab:corr}), in agreement with the conclusions of \citet{Abdo10}.

\begin{table}
\begin{center}
\begin{tabular}{lccc}
\hline
Parameter & XRBs$^1$& Recycled PSs$^2$& $\gamma$-ray flux$^3$\\
\hline
\hline
Pearson r & 0.942 & 0.745 & 0.589\\
$p(>|r|)$ & $4.5\times10^{-5}$ & $0.013$ & 0.124\\
\hline
Spearman r & 0.770 & 0.863 & 0.670\\
$p(>|r|)$ & 0.009 & $0.001$ & 0.068\\
\hline
Kendall $\tau$ & 0.600 & 0.674 & 0.588\\
$p(>|\tau|)$ & 0.016 & $0.006$ & 0.059\\
\hline
\end{tabular}
\end{center}
\caption{Results of statistical tests for correlations of several different measurements of close binaries with our calculations of $\Gamma$.  For all of these tests, the $p$ values show the probability that a correlation arises randomly. Given such low probabilities, there is clear evidence of correlations in all cases.
 1: \citet{Pooley03}, 2: \citet{Bagchi11}, 3: \citet{Abdo10} }
\label{tab:corr}
\end{table}

\section{Conclusions}
In this paper we calculated the stellar interaction rate $\Gamma$ for Galactic globular clusters, directly deprojecting observed surface brightness profiles and then calculating  $\Gamma\propto\int\rho^2/\sigma$. Previous calculations have used simplified relations such as $\Gamma_1\propto\rho_c^2 r_c^3/ \sigma$,  $\Gamma_2\propto\rho_c^{1.5} r_c^2$, or have assumed King-model structures to perform integrations of $\Gamma$ for clusters.  Although our results are generally similar to previous analyses, we find significant differences in several cases, particularly for core-collapsed clusters, which we treat for the first time in the same way as non-core-collapsed clusters.
A major advance in this work is the calculation of uncertainties in our final $\Gamma$ estimates, by using Monte-Carlo simulations to incorporate the effects of observational uncertainties. 

Comparing our $\Gamma$ calculations with observations of close binaries produced by stellar interactions, we found strong evidence for correlations.  This is in agreement with most previous work, but we do find significant differences with key recent results. 
 Comparing our $\Gamma$ to the numbers of XRBs in a GC \citep{Pooley03,Fregeau08}, there is a suggestion that core-collapsed clusters may have \emph{fewer} XRBs than other GCs of similar $\Gamma$, in contrast to \citet{Fregeau08}. Comparing $\Gamma$ to the number of MSPs in GCs, we find extremely strong correlations, in contrast to \citet{Bagchi11}. Finally, we found evidence for correlation of $\Gamma$ with the total $\gamma$-ray fluxes from GCs, in agreement with \citet{Abdo10}.

\acknowledgements 

We thank H. Cohn, P. Lugger, and B. Murphy for helpful discussions.  AB thanks K.S.D. Beach for help with computations.  We acknowledge financial support from  NSERC Discovery Grants (COH, GRS), an Alberta Ingenuity New Faculty Award (COH, JCG), and the Avadh Bhatia Fellowship (JCG). 

\bibliographystyle{apj}
\bibliography{src_ref_list}

\begin{table}[h]
\begin{center}
\begin{tabular}{lccccccccc}
\hline
Name & $ 4 \pi \sigma_c^{-1} \int \rho^2(r) r^2 dr $ & $-\delta$ & $+\delta$ & $\rho_c^2 r_c^3 \sigma_c^{-1}$ &  $-\delta$ & $+\delta$ & $\rho_c^{1.5} r_c^2 $ &  $-\delta$ & $+\delta$ \\
\hline
Terzan 5		&6.80E+3	&		3.02E+3	&		1.04E+3	&		1.86E+3	&		9.34E+2		&	1.99E+3		&	1.40E+3		&	2.85E+2		&	3.23E+2\\
NGC 7078	&4.51E+3		&	9.86E+2	&		1.36E+3	&		5.01E+3	&		2.77E+2		&	3.00E+2		&	6.46E+3		&	1.76E+2		&	1.66E+2\\
NGC 6715	&2.52E+3		&	2.74E+2	&		2.26E+2	&		2.55E+3	&		1.33E+2		&	1.05E+2		&	2.03E+3		&	4.35E+1		&	5.11E+1\\
Terzan 6		&2.47E+3		&	1.72E+3	&		5.07E+3	&		1.78E+3	&		8.51E+2		&	2.34E+3		&	1.30E+3		&	2.41E+2		&	2.79E+2\\
NGC 6441	&2.30E+3		&	6.35E+2	&		9.74E+2	&		2.56E+3	&		1.84E+2		&	1.80E+2		&	3.15E+3		&	1.03E+2		&	1.08E+2\\
NGC 6266	&1.67E+3		&	5.69E+2	&		7.09E+2	&		2.02E+3	&		1.86E+2		&	1.71E+2		&	2.47E+3		&	8.15E+1		&	7.78E+1\\
NGC 1851	&1.53E+3		&	1.86E+2	&		1.98E+2	&		1.54E+3	&		7.66E+1		&	6.98E+1		&	1.91E+3		&	5.41E+1		&	5.48E+1\\
NGC 6440	&1.40E+3		&	4.77E+2	&		6.28E+2	&		1.54E+3	&		5.43E+2		&	1.02E+3		&	1.75E+3		&	1.47E+2		&	1.55E+2\\
NGC 6624	&1.15E+3		&	1.78E+2	&		1.13E+2	&		1.20E+3	&		1.13E+2		&	1.51E+2		&	1.08E+3		&	2.61E+1		&	2.45E+1\\
NGC 6681	&1.04E+3		&	1.92E+2	&		2.67E+2	&		9.81E+2	&		6.04E+1		&	6.59E+1		&	9.64E+2		&	2.75E+1		&	2.63E+1\\
NGC 104		&1.00E+3		&	1.34E+2	&		1.54E+2	&		1.00E+3	&		4.81E+1		&	4.64E+1		&	1.00E+3		&	2.85E+1		&	3.08E+1\\
NGC 5824	&9.84E+2		&	1.55E+2	&		1.71E+2	&		9.16E+2	&		4.15E+1		&	4.89E+1		&	1.22E+3		&	3.18E+1		&	2.98E+1\\
Pal 2			&9.29E+2		&	5.55E+2	&		8.36E+2	&		1.18E+3	&		4.43E+2		&	8.02E+2		&	4.57E+2		&	4.89E+1		&	4.27E+1\\
NGC 2808	&9.23E+2		&	8.27E+1	&		6.72E+1	&		1.15E+3	&		9.79E+1		&	1.11E+2		&	1.21E+3		&	2.49E+1		&	2.67E+1\\
NGC 6388	&8.99E+2		&	2.13E+2	&		2.38E+2	&		9.53E+2	&		7.52E+1		&	7.41E+1		&	1.77E+3		&	4.89E+1		&	4.41E+1\\
NGC 6293	&8.47E+2		&	2.39E+2	&		3.77E+2	&		9.18E+2	&		1.51E+2		&	2.24E+2		&	1.22E+3		&	3.26E+1		&	3.00E+1\\
NGC 362		&7.35E+2		&	1.17E+2	&		1.37E+2	&		8.09E+2	&		3.61E+1		&	3.31E+1		&	5.69E+2		&	1.49E+1		&	1.56E+1\\
NGC 6652	&7.00E+2		&	1.89E+2	&		2.92E+2	&		7.03E+2	&		1.29E+2		&	3.60E+2		&	8.05E+2		&	2.25E+1		&	2.13E+1\\
NGC 6284	&6.66E+2		&	1.05E+2	&		1.22E+2	&		7.50E+2	&		9.82E+1		&	1.22E+2		&	7.97E+2		&	1.95E+1		&	1.76E+1\\
NGC 6626	&6.48E+2		&	9.11E+1	&		8.38E+1	&		6.43E+2	&		1.03E+2		&	1.28E+2		&	6.88E+2		&	1.92E+1		&	1.79E+1\\
\hline\\
\end{tabular}
\end{center}
\caption{$\Gamma$ calculations and 1-$\sigma$ error estimations based on different equations,  all normalized assuming $\Gamma_{NGC104}=1000$. A portion is shown here, the complete table is available in the online edition of the journal.}
\label{tab:gamma_full}
\end{table}

\end{document}